\documentclass[secnumarabic,superscriptaddress,nobibnotes,aps]{revtex4}

\usepackage{amsthm,amsfonts,url}   
\usepackage{bm,amssymb,amsmath,mathrsfs,enumerate,verbatim} 

\theoremstyle{plain}
\newtheorem{thm}{Theorem}[section]

\newtheorem{cor}[thm]{Corollary}

\newtheorem{condition}[thm]{Condition}

\theoremstyle{definition}
\newtheorem{dfn}[thm]{Definition}

\theoremstyle{remark}

\DeclareSymbolFont{AMSb}{U}{msb}{m}{n}
\DeclareMathSymbol{\N}{\mathbin}{AMSb}{"4E}
\DeclareMathSymbol{\Z}{\mathbin}{AMSb}{"5A}
\DeclareMathSymbol{\R}{\mathbin}{AMSb}{"52}
\DeclareMathSymbol{\Q}{\mathbin}{AMSb}{"51}
\DeclareMathSymbol{\I}{\mathbin}{AMSb}{"49}
\DeclareMathSymbol{\C}{\mathbin}{AMSb}{"43}
\DeclareMathSymbol{\F}{\mathbin}{AMSb}{"46}
\DeclareMathSymbol{\E}{\mathbin}{AMSb}{"45}

\DeclareSymbolFont{symbolsC}{U}{txsyc}{m}{n}
\DeclareMathSymbol{\coloneq}{\mathrel}{symbolsC}{66}

\def\tr{\operatorname{tr}}
\def\ket#1{|#1\rangle}
\def\bra#1{\langle#1|}
\def\ketbra#1{| #1 \rangle\langle #1 |}
\def\braket#1#2{\langle#1|#2\rangle}
\def\dket#1{|#1\rangle\!\rangle}
\def\dbraket#1#2{\langle\!\langle#1|#2\rangle\!\rangle}

\def\d{\mathrm{d}}
\def\D{\mathrm{D}}
\def\dd{\partial}
\def\U{\operatorname{U}}
\def\P{\operatorname{P}}
\def\Gr{\operatorname{Gr}}
\def\Span{\operatorname{span}}
\def\rank{\operatorname{rank}}
\def\H{\mathcal{H}}
\def\V{\mathcal{S}}
\def\dmax{s_\mathrm{max}}
\def\dv{s}
\def\c{\mathrm{c}}
\def\psiset{\Psi}

\begin{document}

\title{Generic local distinguishability and completely entangled subspaces}

\author{Jonathan Walgate}
\email{jwalgate@perimeterinstitute.ca}
\affiliation{Perimeter Institute for Theoretical Physics, 31 Caroline St N, Waterloo, Ontario N2L 2Y5, Canada}

\author{A. J. Scott}
\email{andrew.scott@griffith.edu.au} \affiliation{Centre for Quantum Computer Technology and Centre for Quantum Dynamics, 
Griffith University, Brisbane, Queensland 4111, Australia}

\begin{abstract}
A subspace of a multipartite Hilbert space is \emph{completely entangled\/} if it contains no product states. Such subspaces can be
large with a known maximum size, $\dmax$, approaching the full dimension of the system, $D$. We show that almost all subspaces with
dimension $\dv\leq\dmax$ are completely entangled
and then use this fact to prove that $n$ random pure quantum states
are unambiguously locally distinguishable if and only if $n \leq D - \dmax$. This condition holds for almost all sets of states
of all multipartite systems, and reveals something surprising. The criterion is identical for separable and for nonseparable
states: entanglement makes no difference.
\end{abstract}

\maketitle

\section{Introduction}

This paper addresses two broad questions. What properties characterize subspaces of multipartite quantum systems? How difficult
is it to locally discriminate quantum states? These questions are fundamental to quantum information theory, but remarkably hard
to answer in full. The sheer variety of quantum states makes a complete generalization difficult, and this is especially clear in
multipartite Hilbert spaces, which confront us with a multitude of incompatible kinds of entanglement~\cite{measure1,measure2,
measure3}. Amid such diversity, where exceptions abound to every rule of thumb, how can we form any intuition about typical
quantum properties? Do typical properties even exist?

Yes, and this is thanks to the concentration of measure phenomenon expressed in L\'{e}vy's Lemma~\cite{Levy}. Loosely put, in
Hilbert spaces of high dimension, pure quantum states chosen uniformly at random are likely to have average properties with
probability approaching one. A property possessed by a random state with high probability might be regarded as typical; a
property possessed by a random state with unit probability, that is, by almost all states, is regarded as generic. When we
characterize random states, and random subspaces, we are thus studying properties of which that can be regarded as typical or
generic for all. In this article we study generic properties. We will sketch some relevant facts concerning random states and
subspaces in Sec.~\ref{randomness}, but for a thorough discussion, consider Refs.~\cite{random1,random2,Hayden1,Montanaro} and
the references therein.

In Sec.~\ref{Random Subspaces} we provide a complete answer to an important question: when are random subspaces of a multipartite
quantum system completely entangled? That is, when are they void of product states? The answer: whenever any one subspace has
this property. The same is found true for random subspaces of a bipartite system in relation to states of a given Schmidt rank.
Highly entangled subspaces are relevant to quantum error-correcting codes~\cite{Scott,Gour}. Random subspaces have been studied
in a wide variety of contexts, such as entanglement measures \cite{Hayden1}, superdense coding \cite{Abeyes}, and enhancing the
capacity of private quantum channels \cite{Spekkens}. This list of applications is unlikely to be exhaustive; as we shall show,
completely entangled subspaces are ubiquitous.

Three previous results are directly relevant to this paper. Hayden, Leung and Winter~\cite{Hayden1} have given a sufficient
condition on the dimension of a random subspace of a bipartite system to typically contain only highly entangled states.
Although only a small fraction of the overall Hilbert space is spanned, the subspace dimension can be made
surprisingly large relative to the dimension of the subsystems. Wallach~\cite{Wallach} and 
Parthasarathy~\cite{Parthasarathy} have independently proven that in a multipartite Hilbert space, $\H=\bigotimes_{j} \H_j$, the maximum
dimension of a completely entangled subspace is $\dmax = D - \sum_{j} (d_{j}-1) -1$, where $d_j=\dim\H_j$
and $D=\dim\H=\prod_j d_j$. Parthasarathy also provides an explicit construction. Recently, Cubitt, Montanaro and
Winter~\cite{Cubitt} have extended this result in the bipartite case and obtained maximal constructions of subspaces
that are void of states with low Schmidt rank. We show that almost all subspaces have the same properties as these 
maximal constructions. Thus, for example, almost all multipartite subspaces with dimension $\dv \leq \dmax $ contain
no product states.

In Sec.~\ref{Random States} we answer a question intimately related to the above: when is a set of $n$ random pure states unambiguously locally
distinguishable? The answer: whenever $n \leq D-\dmax = 1 + \sum_{j} (d_{j}-1)$. State discrimination is perhaps the
most elementary physical task of all, and studying it under a restriction to local operations and classical communication (LOCC)
provides insight into the local structure of quantum information. It has attracted much interest in recent years, but progress has
been slowed by the abundance of special cases lurking in Hilbert space. Results have been restricted mostly to low dimensional
systems, specific multipartite structures, and small numbers of states~\cite{walgate,virmani,ghosh,duan2}. These limitations are
overcome for random states, which allow us to provide a complete generic solution for the unambiguous local distinguishability of
pure states, over all multipartite quantum systems. A surprising fact then emerges: in almost all cases,
entanglement makes no difference as to whether the states can be distinguished or not. Sets of random
product states obey the exact same criterion as sets of random highly entangled states.

Our result recalls a recent discovery of Duan \emph{et al}.~\cite{Duan} that every complete basis for a multipartite Hilbert space contains
some subset of exactly $1+\sum_j(d_{j}-1)$ locally unambiguously distinguishable states. One consequence of our
result is that for almost all bases this is maximal. We also extend our analysis to the case of multiple copies, which allows
some insight into the difficulties typically imposed by the restriction that separate measurements be performed on parts of
a quantum system. To obtain the same information, separable measurements consume exponentially more copies of a system than are
needed globally.

Finally, in Sec.~\ref{open} we discuss some observations and open questions. 

\section{Randomness} \label{randomness}

Throughout this article, we will use the adjectives `random' and `generic', and the adjectival phrases `almost surely' and
`almost all', to describe pure quantum states and subspaces. By this we will mean the following.

Let $\H=\C^D$ be a Hilbert space of dimension $D$. A pure state $\psi$ of $\H$ is represented by a unit vector $\ket{\psi}\in\H$,
modulo a phase factor, $e^{i\theta}\ket{\psi}\sim\ket{\psi}$, and is thus a member of projective Hilbert space: $\psi\in\P(\H)$.
This is just the space of lines passing through the origin in $\H$, or equivalently, the space of one-dimensional subspaces of
$\H$. The space of $s$-dimensional subspaces of $\H$ is the Grassmannian $\Gr_{\dv}(\H)$. Thus $\P(\H)\cong\Gr_1(\H)$.

There is a natural uniform probability measure on $\Gr_{\dv}(\H)$ induced by the unit Haar measure $\nu$~\cite{Haar} on the unitary group
$\U(D)$. Explicitly, we fix a subspace $\V\in\Gr_{\dv}(\H)$ and define $\mu_{\dv}(E)\coloneq\nu(\{U\in\U(D):U\V\in E\})$
for any (Borel) subset $E\subseteq\Gr_{\dv}(\H)$. This is the unique unitarily invariant probability
measure on $\Gr_{\dv}(\H)$. The probability measure induced on $\P(\H)$ in this way will be denoted by $\mu$. We are now in a
position to define random states and subspaces.

\begin{dfn}
A \emph{random pure state\/} $\psi$ of the Hilbert space $\H$ is a random variable that takes values in $\P(\H)$ and is
distributed according to the uniform probability measure $\mu$ on $\P(\H)$.
\end{dfn}

\begin{dfn}
A \emph{random ($s$-dimensional) subspace\/} $\V$ of the Hilbert space $\H$ is a random variable that takes values in $\Gr_{\dv}(\H)$
and is distributed according to the uniform probability measure $\mu_{\dv}$ on $\Gr_{\dv}(\H)$.
\end{dfn}

A random subspace (or random pure state) can thus be generated from a fixed subspace (of the same dimension), by applying a unitary chosen at
random according to the Haar measure. A random pure state of a multipartite Hilbert space will, with probability one,
have maximum Schmidt rank, and typically be highly entangled. A set of $s\leq D$ random pure states will, with probability one, be linearly independent.
The span of these states is then a random $s$-dimensional subspace. This is because the measure induced on $\Gr_{\dv}(\H)$ is unitarily
invariant, and hence, can only be $\mu_{\dv}$. If $\V\leq\H$ is a random $s$-dimensional subspace, then
$\V^{\perp}$ is a random $(D - s)$-dimensional random subspace, where $\V\oplus\V^{\perp}=\H$.

We will occasionally use the terms `generic', `almost surely' and `almost all'. When we say a random pure state {\em almost surely\/}
has some specific property, then we mean that this property occurs with unit probability. This means the property is true for all pure states
except those on a set of zero $\mu$-measure. Thus, put simply, {\em almost all\/} pure states have this property; more
precisely, {\em $\mu$-almost all\/} pure states have this property. The property is then called {\em generic\/}. The same
terminology is used for random subspaces and sets of random pure states, except that $\mu$ is replaced by $\mu_{\dv}$ or the
product measure $\mu\times\dots\times\mu$, respectively.

\section{Completely Entangled Random Subspaces}\label{Random Subspaces}

It is known that reasonably large subspaces of Hilbert spaces can be completely entangled. A dramatic demonstration of this fact
was provided by Hayden \emph{et al}. \cite{Hayden1}. Let $\H = \H_1 \otimes \H_2$
with $d_2\geq d_1\geq 3$, where $d_1=\dim\H_1$, $d_2=\dim\H_2$,
and we set $D=d_1d_2=\dim\H$. Then for any $0 < c < 1$ there exists a subspace $\V\leq\H$ of dimension
\begin{equation}\label{haydenbound}
\dv \;=\;  \left\lfloor  \frac{c^{2.5}}{1753}\,D\right\rfloor \;,
\end{equation}
containing only states with entanglement $E\geq (1-c)\log d_1 - (1/\ln 2)d_1/d_2$ bits, where
$E(\psi)\coloneq -\tr\rho_\mathrm{A}\log\rho_\mathrm{A}$ with $\rho_\mathrm{A}=\tr_\mathrm{B}\ketbra{\psi}$. Moreover,
the probability for a random subspace of dimension $\dv$ to not have this property is bounded above by
\begin{equation}
\left(\frac{15}{c}\right)^{2\dv}\exp\left(-\,\frac{(D-1)c^2}{32\pi^2\ln(2)}\right)\;.
\end{equation}
Thus, in increasing dimensions, random subspaces of dimension $\dv$ (or smaller) are typically completely entangled, with
probability exponentially approaching one. These subspaces are surprisingly large. For instance, if
$\H_1$ and $\H_2$ are both $n$ qubit systems, then $\V$ is a subspace of dimension equal to
that of a $2n - O(1)$ qubit system. Nevertheless, $\dv$ is always a small fraction of $D$.

Wallach~\cite{Wallach} and Parthasarathy~\cite{Parthasarathy} independently showed that much larger dimensions could be 
spanned without encompassing any product states. In particular, the following was proven.

\begin{thm}[Wallach~\cite{Wallach}, Parthasarathy~\cite{Parthasarathy}]\label{Wallach}
Let\/ $\H=\bigotimes_j \H_j$ be a multipartite Hilbert space with\/ $\dim\H_j=d_j$ and\/ $\dim\H= D =\prod_{j} d_j$.
Then there exists a completely entangled subspace\/ $\V\leq\H$ of dimension\/ $\dv$ if and only if
\begin{equation}
\dv \;\leq\; D - \sum_{j} (d_{j}-1)-1 \;.
\end{equation}
\end{thm}

We thus define the maximum possible dimension of a completely entangled subspace:
\begin{equation}
\dmax \;\coloneq\; D - \sum_{j} (d_{j}-1) -1 \;.
\end{equation}
In comparison to the result of Hayden {\em et al.} for bipartite systems (\ref{haydenbound}), if
$\H=\H_1\otimes\H_2$, where $\H_1$ and $\H_2$ are both $n$ qubit systems,
then a completely entangled subspace can have a dimension equal to that of a $2n - O(2^{-n})$ qubit system.

We extend Theorem~\ref{Wallach} in Corollary~\ref{subspace2} below, where we show that a random subspace of dimension $\dmax$ or
less is almost surely void of product states. Two different but equivalent approaches will be taken. First, we use a theorem of
differential geometry to give an explicit, straightforward proof. Then a refined approach is taken, which is based in the theory
of algebraic geometry. While the latter method is arguably more obscure, given the conceptual hurdles that must be overcome, it
will ultimately prove more powerful.

\subsection{Completely entangled random subspaces: A differential geometry approach}

Consider the following variation of Sard's theorem of differential geometry~\cite{Abraham67,Hirsch76,Abraham83}.
A proof was sketched by Chow {\it et al.}~\cite{Chow}, but consider Keerthi {\it et al.}~\cite{Keerthi} for a guide
to its use in establishing genericity. Let $f:X\subseteq\R^M\rightarrow\R^N$ be $\mathcal{C}^1$, i.e. a differentiable function whose first derivative is continuous.
We say that $x\in X$ is a {\em regular point\/} of $f$ if the Jacobian $\D f(x)$ has full row rank. If the set
$f^{-1}(y)$, the preimage of $y\in \R^N$, contains only regular points, then $y$ is called a {\em regular value\/} of $f$.
The phrase `almost all' is used in the sense of Lebesgue measure in the following.

\begin{thm}[Parametrized Sard Theorem]\label{thm:sard}
Let\/ $X\subseteq \R^L$ and\/ $Y\subseteq \R^M$ be open, and let\/ $f:X\times Y \rightarrow \R^N$ be\/ $\mathcal{C}^r$ with\/
$r>\max\{0,L-N\}$. If\/ $0\in\R^N$ is a regular value of\/ $f$, then for almost all\/ $y\in Y$,\/
$0$ is a regular value of\/ $f(\,\cdot\,,y)$. In particular,\/ $\tilde{X}(y)\coloneq\{x\in X: f(x,y)=0\}$ is either empty or a differentiable manifold of dimension\/ $L-N$, for
almost all\/ $y\in Y$.
\end{thm}
\begin{cor}\label{cor:sard}
Let\/ $X\subseteq \R^L$ and\/ $Y\subseteq \R^M$ be open, and let\/ $f:X\times Y \rightarrow \R^N$ be\/ $\mathcal{C}^r$ with\/
$r>\max\{0,L-N\}$. If\/ $0\in\R^N$ is a regular value of\/ $f$ and\/ $L<N$, then\/ $\tilde{X}(y)\coloneq\{x\in X: f(x,y)=0\}$
is empty for almost all\/ $y\in Y$.
\end{cor}

The Parametrized Sard Theorem, or more specifically Corollary~\ref{cor:sard}, is applied to the current problem by simply
associating the variable $x$ with an arbitrary product pure state, the variable $y$ with an arbitrary set of $n$ general pure
states, and then defining $f$ as the vector function of inner products between the product and general states. The trick is to
choose $x$ and $y$ in such a way that $0$ is guaranteed a regular value of $f$. The fact that this is so easily done is a feature
of complex projective space. The problem then reduces to a parameter counting argument.

\begin{thm} \label{subspace1}
Let\/ $\H=\bigotimes_j \H_j$ be a multipartite Hilbert space with\/ $\dim\H_j=d_j$.
Then for $\mu$-almost all pure states\/ $\psi_1,\dots,\psi_n\in\P(\H)$, if
\begin{equation}
n \;>\; \sum_j (d_j-1) \;,
\end{equation}
no product state is orthogonal to all of\/ $\ket{\psi_1},\dots,\ket{\psi_n}$.
\end{thm}

\begin{proof}
Let $\{ \ket{k_1}\otimes\dots\otimes\ket{k_m}\}_{k_{j} = 1,\dots,d_{j}}$ be an orthonormal product basis for
$\H= \bigotimes_{j=1}^m \H_j$. Parametrize an unnormalized product pure state of $\H$ as
\begin{equation}\label{productparams}
\dket{\xi(a^1,\dots,a^m)} \;\coloneq\; \sum_{k_1,\dots,k_m}\big(a^1_{k_1}\big)^* \dots
\big(a^m_{k_m}\big)^*\,\ket{k_1}\otimes\dots\otimes\ket{k_m} \;,
\end{equation}
where each $a^j_{k_j}\in\C$, and thus, $a^j\in\C^{d_j}$. The notation $\dket{\,\cdot\,}$ will be used for the duration of
this proof to emphasize that such vectors are generally unnormalized. Parametrize an unnormalized general pure state of $\H$ as
\begin{equation}
\dket{\eta(b)} \;\coloneq\; \sum_{k_1,\dots,k_m} b_{k_1 \dots k_m}\,\ket{k_1}\otimes\dots\otimes\ket{k_m} \;,
\end{equation}
where each $b_{k_1 \dots k_m}\in\C$, and thus, $b\in\C^D$.
Furthermore, define the polynomial functions
\begin{align}
g_p(a^1,\dots,a^m;b^1,\dots,b^n) &\;\coloneq\; \dbraket{\xi(a^1,\dots,a^m)}{\eta(b^p)} \\
&\;=\; \sum_{k_1,\dots,k_m}  {a^1_{k_1}} \cdots {a^m_{k_m}} b_{k_1 \dots k_m}^p \;,
\end{align}
for $p=1,\dots,n$. Thus there is a product state orthogonal to the subspace
\begin{equation}
\V(b^1,\dots,b^n) \;\coloneq\; \Span\big\{\dket{\eta(b^1)},\dots,\dket{\eta(b^n)}\big\}
\end{equation}
if and only if $g_1(a^1,\dots,a^m;b^1,\dots,b^n)=\dots=g_n(a^1,\dots,a^m;b^1,\dots,b^n)=0$ has a
solution in the variables $a^1,\dots,a^m$.

Now define the sets
\begin{equation}
A_{k_1\dots k_m}\;\coloneq\;\big\{(a^1,\dots,a^m)\in\C^{d_1}\times\dots\times\C^{d_m}:a^1_{k_1}=\dots=a^m_{k_m}=1\big\}\;,
\end{equation}
which specify subsets of product states,
\begin{equation}
\Xi_{k_1\dots k_m} \;\coloneq\; \big\{\dket{\xi(a^1,\dots,a^m)}:(a^1,\dots,a^m)\in A_{k_1\dots k_m}\big\} \;,
\end{equation}
whose union, $\Xi\coloneq\cup_{k_1\dots k_m}\Xi_{k_1\dots k_m}$, includes a representative of each normalized product state,
i.e. every normalized product state can be written as $\ket{\phi}=c\dket{\xi}$ for some $\dket{\xi}\in\Xi$ and $c\in\C$.

First consider $\V^{\perp}\cap \Xi_{1 \ldots 1}$. Since
\begin{equation}
\frac{\dd g_p}{\dd b_{1\ldots 1}^q} \;=\; \delta_{qp}\, a^1_1\cdots a^m_1 \;=\; \delta_{qp} \;,
\end{equation}
if $(a^1,\dots,a^m)\in A_{1\ldots 1}$, the Jacobian $\D g(a^1,\dots,a^m;b^1,\dots,b^n)$ has full
(complex) rank for all $(a^1,\dots,a^m)\in A_{1\ldots 1}$ and all $b^1,\dots,b^n$. Now set
\begin{equation}
f(x,y)\;\coloneq\; \big(\Re(g_1),\dots,\Re(g_n),\Im(g_1),\dots\big)\;,
\end{equation}
where
\begin{align}
x &\;\coloneq\;\big(\Re(a^1_2),\dots,\Re(a^1_{d_1}),\Re(a^2_2),\dots,\Re(a^m_{d_m}),\Im(a^1_2),\dots\big)\;\in\;\R^{2(d_1+\cdots+d_m-m)}\;,  \\
y &\;\coloneq\;\big(\Re(b_{1\ldots 1}^1),\dots,\Re(b_{d_1\ldots d_m}^1),\Re(b_{1\ldots 1}^2),\dots,\Re(b_{d_1\ldots
d_m}^n),\Im(b_{1\ldots 1}^1),\dots\big)\;\in\;\R^{2nD}\;.
\end{align}
Since each $g_p$ is an analytic function in its variables, the Jacobian $\D f(x,y)$ will also have full (real) rank for all $x$
and $y$ (consider the Cauchy-Riemann equations). Now with $X=\R^L$, $Y=\R^M$, $L=2(d_1+\cdots+d_m-m)$, $M=2nD$ and $N=2n$, by
Corollary~\ref{cor:sard}, $\tilde{X}(y)\coloneq\{x\in X: f(x,y)=0\}=\emptyset$ for almost all $y\in Y$ (in the sense
of Lebesgue measure) whenever $L<N$. Equivalently, $\V^{\perp}(y)\cap \Xi_{1 \ldots 1}=\emptyset$ for almost all $y\in Y$ whenever $n> \sum_{j} (d_{j}-1)$.

Repeating the above argument for all sets $A_{k_1\dots k_m}$, and given that the finite
union of sets of measure zero is again of measure zero, we conclude that $\V^{\perp}(y)\cap\Xi=\emptyset$ for almost
all $y\in Y$ whenever $n> \sum_{j} (d_{j}-1)$.

Finally, note that we can rewrite each $\dket{\eta(b^j)}=r_je^{i\theta_j}\ket{\psi_j}$, where $r_j\geq 0$, $0\leq\theta_j<2\pi$
and $\ket{\psi_j}$ is a normalized pure state specified by $\psi_j\in\P(\H)$. Now since the current measure on
unnormalized pure states (essentially the Lebesgue measure on $\R^{2D}$) is related to the uniform probability measure $\mu$ on $\P(\H)$ (described in Sec.~\ref{randomness})
through the differential
\begin{equation}
 \d\Re(b_{1\ldots 1})\dots\d\Re(b_{d_1\ldots d_m})\d\Im(b_{1\ldots 1})\dots\d\Im(b_{d_1\ldots d_m}) \;=\; r^{2D-1}\d r\d\theta\d\mu(\psi) \;,
\end{equation}
and furthermore, $\V(b^1,\dots,b^n)=\Span\{\ket{\psi_1},\dots,\ket{\psi_n}\}$, independent of each $r_j$
and $\theta_j$, the above will be contradicted unless $\V^\perp$ contains no product states for $\mu$-almost all
$\psi_1,\dots,\psi_n\in\P(\H)$, whenever $n>\sum_{j} (d_{j}-1)$.
\end{proof}

Recall that the product measure $\mu\times\dots\times\mu$ on $\P(\H)\times\dots\times\P(\H)$ (both $n$ times) induces
the uniform probability measure $\mu_{n}$ on $\Gr_{n}(\H)$, since the former is unitarily invariant,
and $n\leq D$ random pure states almost surely span $n$ dimensions.
The following is thus equivalent to Theorem~\ref{subspace1}.

\begin{cor}\label{subspace2}
Let\/ $\H=\bigotimes_j \H_j$ be a multipartite Hilbert space with\/ $\dim\H_j=d_j$ and\/ $\dim\H= D =\prod_{j} d_j$.
Then for\/ $\mu_{\dv}$-almost all subspaces\/ $\V\in\Gr_{\dv}(\H)$, if
\begin{equation}\label{bound}
\dv \; \leq \; \dmax \;=\; D - \sum_j (d_j-1) -1 \;,
\end{equation}
$\V$ contains no product states.
\end{cor}

Random subspaces do not contain product states unless they \emph{have\/} to, since if $\dv > \dmax$, then by
Theorem~\ref{Wallach}, {\em all\/} subspaces $\V\in\Gr_{\dv}(\H)$ contain at least one product state.
Wallach and Parthasarathy both proved this last fact using algebraic geometry. In fact, all of the above results are consequences
of this theory, which we will explain next. We end this subsection with a straightforward corollary of
Theorem~\ref{subspace1}. This will be needed later in the article.

\begin{cor} \label{subspace4}
Let\/ $\H=\bigotimes_j \H_j$ be a multipartite Hilbert space with\/ $\dim\H_j=d_j$, and let\/ $c$ be any positive integer.
Then for $\mu$-almost all pure states\/ $\psi_1,\dots,\psi_n\in\P(\H)$, if
\begin{equation}
n \;>\; c\sum_j (d_j-1) \;,
\end{equation}
no product state of\/ $\H^{\otimes c}=(\bigotimes_j \H_j)^{\otimes c}$ (between all subsystems) is orthogonal to all of\/ $\ket{\psi_1}^{\otimes c},\dots,\ket{\psi_n}^{\otimes c}$.
\end{cor}

\begin{proof}
Let $n>c\sum_j (d_j-1)$ and let $\phi$ be a product state of $\H^{\otimes c}$,
\begin{equation}
\ket{\phi}\;=\;\ket{\phi_1}\otimes\cdots\otimes\ket{\phi_c}\;,
\end{equation}
where each $\phi_j$ is a product state of $\H$. Suppose that $\bra{\phi}(\ket{\psi_k}^{\otimes c})=0$ for all $k$, i.e.,
\begin{equation}
\braket{\phi_1}{\psi_k}\braket{\phi_2}{\psi_k}\cdots\braket{\phi_c}{\psi_k}\;=\;0\;,
\end{equation}
for $k=1,\dots,n$. To satisfy all $n$ of these equations there must be at least one product state,
$\phi_{j'}$ say, with $\braket{\phi_{j'}}{\psi_k}=0$ for $m>\sum_j (d_j-1)$ values of $k$. But if
the states $\psi_k$ are chosen randomly, this is almost surely impossible, since otherwise we would contradict
Theorem~\ref{subspace1}. Thus there is no product state orthogonal to all of
$\ket{\psi_1}^{\otimes c},\dots,\ket{\psi_n}^{\otimes c}$ for $\mu$-almost all choices of\/ $\psi_1,\dots,\psi_n\in\P(\H)$.
\end{proof}

\subsection{Completely entangled random subspaces: An algebraic geometry approach}

A {\em complex projective variety\/} $X$ is an algebraic subset of $\P(\C^D)$, i.e., $X$ can be specified as
the common zero locus of a set $S\subset\C[x_1,\dots,x_D]$ (which can be assumed finite) of homogeneous polynomials
in the coordinates of $\C^D$,
\begin{equation}
X \;=\; V(S) \;\coloneq\; \big\{x\in\P(\C^D) : f(x_1,\dots,x_D)=0\text{ for all }f\in S \big\} \;,
\end{equation}
where $x_k\coloneq\braket{k}{x}$ for some fixed basis $\{\ket{k}\}_{k=1,\dots,D}$ of $\C^D$. This makes sense because a
homogeneous polynomial satisfies $f(cx_1,\dots,cx_D)=c^kf(x_1,\dots,x_D)$ by definition, and thus, if $f$ vanishes on any point
of the line specified by $x$, then it vanishes for all. An example of a variety is a projective subspace, or {\em
$(\dv-1)$-plane\/}:
\begin{equation}
\Lambda_\V \;\coloneq\; \P(\V) \;=\; \big\{x\in\P(\C^D) : \ket{x}\in\V \big\} \;,
\end{equation}
where $\V\in\Gr_\dv(\C^D)$. This is of course the locus of solutions to $D-\dv$ homogeneous linear equations:
$\V^\perp\ket{x}=0$. In general, complex projective varieties are finite collections of submanifolds of
$\P(\C^D)$. There are many good texts on algebraic geometry, of which, Harris~\cite{Harris} is perhaps the
most readable introduction. We should remark, however, that some authors reserve the term `variety'
for what we will be calling an irreducible variety; we follow Harris and allow any algebraic set.

Note that the intersection of any two varieties is always another variety, since
$V(S)\cap V(T)=V(S\cup T)$. In fact, because $\C[x_1,\dots,x_D]$ is a Noetherian ring, we can show
that the intersection of any number of varieties is another. Additionally, the union of any finite
number of varieties is another variety, since for any two, $V(S)\cup V(T)=V(ST)$, where $ST=\{fg:f\in S,g\in T\}$.
The empty set and entire space are also varieties: $\emptyset=V(\{1\})$ and $\P(\C^D)=V(\emptyset)$. These
properties define a topology on $\P(\C^D)$, called the {\em Zariski topology\/}, where the subvarieties
$X\subseteq\P(\C^D)$ are defined as the closed sets, or more formally, the {\em Zariski closed sets\/}. A
{\em Zariski open set\/} $X^\c$ is then the complement in
$\P(\C^D)$ of some variety $X$. We should remark here, however, that the Zariski topology is only a formal
construct of algebraic geometry and reflects little of the usual topology of complex projective space. Any
nonempty Zariski open set is both of full $\mu$-measure and dense in the usual topology, 
being the complement of a finite collection of proper submanifolds of $\P(\C^D)$.

A variety $X$ is called {\em irreducible\/} if for each pair of subvarieties $Y,Z\subseteq X$ such that
$Y\cup Z=X$, either $Y=X$ or $Z=X$. Any $(\dv-1)$-plane, for example, from a single point up to the entire space itself,
is an irreducible variety. The concept of dimension can be defined for
irreducible varieties. There are various equivalent definitions, three of which are as follows.

\begin{dfn}[Harris~{\cite[p.~134]{Harris}}]\label{dfn:dim}
The {\em dimension\/} of an irreducible complex projective variety $X\subseteq\P(\C^D)$, denoted
$\dim X$, can be defined in three equivalent ways:
\begin{enumerate}
\item $\dim X$ is the smallest integer $d$ such that there exists a $(D-d-2)$-plane disjoint from $X$.
\item $\dim X$ is the smallest integer $d$ such that the general $(D-d-2)$-plane is disjoint from $X$.
\item $\dim X$ is that integer $d$ such that the general $(D-d-1)$-plane intersects $X$ in a finite set of points.
\end{enumerate}
\end{dfn}

This means $\dim\Lambda_\V=\dim\V-1$, and in particular, $\dim\P(\C^D)=D-1$. The equivalence
of these definitions should be shown, but we will take them as fact. All three indirectly imply
something important about the number of product pure states contained in a subspace of multipartite Hilbert space.

Consider Definition~\ref{dfn:dim}.1 for example. Define $\sigma:\P(\C^{d_1})\times\P(\C^{d_2})\rightarrow\P(\C^{d_1d_2})$ by $\sigma(x,y)=z$ where $
z_{jk}=x_jy_k$. The image of this map $\Sigma_{d_1,d_2}\coloneq\sigma(\P(\C^{d_1})\times\P(\C^{d_2}))$ is then a variety, called the {\em Segre variety\/}.
It is the locus of solutions to the quadratic equations $z_{ij}z_{kl}-z_{il}z_{kj}=0$. Importantly, $\sigma$ defines
the {\em Segre embedding\/} of the Cartesian product of two projective spaces into a larger projective space.
It shows how the Cartesian product of two varieties, $X\times Y\subseteq\P(\C^{d_1})\times\P(\C^{d_2})$, can be considered
another variety in the larger space, $\sigma(X\times Y)\subseteq\P(\C^{d_1d_2})$.
If $X$ and $Y$ are irreducible, then so is $\sigma(X\times Y)$, with $\dim \sigma(X\times Y)=\dim X + \dim Y$.
In particular, $\Sigma_{d_1,d_2}$ is irreducible with dimension
\begin{equation}
\dim \Sigma_{d_1,d_2}\;=\;\dim\P(\C^{d_1})+\dim\P(\C^{d_2})\;=\; (d_1-1)+(d_2-1) \;.
\end{equation}
Now define the general Segre variety $\Sigma_{d_1,\dots,d_m}\coloneq \sigma(\Sigma_{d_1,\dots,d_{m-1}}\times\P(\C^{d_m}))$ recursively.
Then
\begin{equation}\label{segredim}
\dim\Sigma_{d_1,\dots,d_m} \;=\; \sum_j(d_j-1) \;.
\end{equation}
Finally, given that $\Sigma_{d_1,\dots,d_m}\subseteq\P(\C^D)$, $D=\prod_j d_j$, is really just
the set of product states for the multipartite Hilbert space $\H=\bigotimes_j\C^{d_j}$,
Theorem~\ref{Wallach} is a straightforward consequence of Definition~\ref{dfn:dim}.1.

Now consider Definition~\ref{dfn:dim}.2. We first need to describe what is meant by a `general' $(\dv-1)$-plane.
An algebraic subset of the Grassmannian $\Gr_{\dv}(\C^D)$ can be viewed as a projective variety under the
Pl\"{u}cker embedding. We refer to Harris~\cite{Harris} for the details, but for our purposes, simply note that
there is a Zariski topology on $\Gr_{\dv}(\C^D)$ analogous to that for $\P(\C^D)$. Now when we say that
the {\em general\/} $(\dv-1)$-plane has some property, we mean that the subset of $(\dv-1)$-planes with this property,
$\{\Lambda_\V\}_{\V\in S}$, is specified by a parameter set $S\subseteq\Gr_{\dv}(\C^D)$ that contains a nonempty
Zariski open subset of $\Gr_{\dv}(\C^D)$. This means $S$ has full $\mu_\dv$-measure. Thus,
given the dimension of the Segre variety~(\ref{segredim}), Corollary~\ref{subspace2} (and Theorem~\ref{subspace1}) 
becomes a straightforward consequence of Definition~\ref{dfn:dim}.2.

Finally, consider the last definition of dimension, Definition~\ref{dfn:dim}.3. As implied by this variation,
for any irreducible variety $X$, the general $(D-\dim X-1)$-plane intersects $X$ at a finite nonzero number of points.
This number is the degree of the variety.

\begin{dfn}[Harris~{\cite[p.~225]{Harris}}]\label{dfn:deg}
The {\em degree\/} of an irreducible complex projective variety $X\subseteq\P(\C^D)$, denoted
$\deg X$, is the number of points at which the general $(D-\dim X-1)$-plane intersects $X$.
\end{dfn}

It is in fact a feature of complex projective space that any two irreducible varieties, $X,Y\subseteq\P(\C^D)$,
must intersect whenever $\dim X +\dim Y \geq D-1$ (Theorem~\ref{Wallach} is partly based on this fact).
In the case of the Segre variety,
\begin{equation}
\deg\Sigma_{d_1,\dots,d_m} \;=\; \frac{\big(\sum_j (d_j-1) \big)!}{\prod_j (d_j-1)!}\;,
\end{equation}
which is derived recursively with the help of Harris~\cite[Ex.~19.2]{Harris}. The following now generalizes Corollary~\ref{subspace2}.

\begin{cor}\label{subspace3}
Let\/ $\H=\bigotimes_j \H_j$ be a multipartite Hilbert space with\/ $\dim\H_j=d_j$ and\/ $\dim\H= D =\prod_{j} d_j$.
Then for\/ $\mu_\dv$-almost all subspaces\/ $\V\in\Gr_{\dv}(\H)$, the number of product states contained in\/ $\V$ is exactly
\begin{equation}
\begin{cases}
0\,, & \text{if\/ $\,\dv\leq\dmax = D - \sum_j (d_j-1) -1$}\,;\\
\frac{(\sum_j (d_j-1) )!}{\prod_j (d_j-1)!}\,, & \text{if\/ $\,\dv = \dmax +1 $}\,; \\
\infty\,, & \text{otherwise}\,.
\end{cases}
\end{equation}
\end{cor}

\subsection{Random subspaces void of states with low Schmidt rank}

Let $\H=\H_1\otimes\H_2$ be a bipartite Hilbert space of dimension $D=d_1d_2$, where $\dim\H_j=d_j$.
A pure state $\ket{\psi}\in\H$ is said to have {\em Schmidt rank\/} $r$ if the
$d_1\times d_2$ matrix $M(\psi)$ with components $[M(\psi)]_{jk}=(\bra{j}\otimes\bra{k})\ket{\psi}$ has
rank $r$. This means the Schmidt decomposition of $\ket{\psi}$, defined by the singular value
decomposition of $M(\psi)$, has $r$ terms:
\begin{equation}
\ket{\psi} \;=\; \sum_{l=1}^r \sqrt{\lambda_l}\ket{u_l}\otimes\ket{v_l}\;,
\end{equation}
given the singular value decomposition $[M(\psi)]_{jk}=\sum_{l=1}^r \sqrt{\lambda_l}\braket{j}{u_l}\braket{{v_l}^*}{k}$
(conjugation in the standard basis). All bipartite entanglement measures are functions of the Schmidt coefficients $\lambda_l$,
one of which, is the Schmidt rank.

Theorem~\ref{Wallach} gives the maximum dimension of a subspace $\V\leq\H$
containing no states of Schmidt rank $1$.
What about subspaces void of states with Schmidt rank $r$ or less? This scenario was studied by Cubitt,
Montanaro and Winter~\cite{Cubitt}. We will now briefly mention what algebraic geometry implies.

The {\em determinantal variety} is defined as
\begin{equation}
M_r \;\coloneq\; \left\{x\in\P(\C^{d_1d_2}) : \rank M(x) \leq r \right\} \;,
\end{equation}
which is the common zero locus of the $(r+1)\times(r+1)$ minor determinants of $M(x)$. Note that $M_1=\Sigma_{d_1,d_2}$,
the Segre variety. In general, $M_r$ is the set of states with Schmidt rank $r$ or less. From Harris~\cite[pp. 151 and 243]{Harris},
\begin{equation}
\dim M_r \;=\; d_1d_2-(d_1-r)(d_2-r)-1 \;,
\end{equation}
and for $d_2\geq d_1> r $,
\begin{equation}
\deg M_r \;=\; \prod_{j=0}^{d_1-r-1} \frac{(d_2+j)!j!}{(r+j)!(d_2-r+j)!}\;.
\end{equation}
The following can thus be deduced.

\begin{cor}[Cubitt~\emph{et al}.~\cite{Cubitt}]
Let\/ $\H=\H_1\otimes\H_2$ be a bipartite Hilbert space with\/ $\dim\H_j=d_j$.
Then there exists a subspace\/ $\V\leq\H$ of dimension\/ $\dv$, completely void of states with Schmidt rank $r$ or less, if and only if
\begin{equation}
\dv \;\leq\; (d_1-r)(d_2-r) \;.
\end{equation}
\end{cor}

\begin{cor}\label{Schmidt}
Let\/ $\H=\H_1\otimes\H_2$ be a bipartite Hilbert space with\/ $\dim\H_j=d_j$, where\/ $d_2\geq d_1$.
Then for\/ $\mu_\dv$-almost all subspaces\/ $\V\in\Gr_{\dv}(\H)$, the number of states with Schmidt rank $r$ or less
contained in\/ $\V$ is exactly
\begin{equation}
\begin{cases}
0\,, & \text{if\/ $\,\dv\leq \dmax' =(d_1-r)(d_2-r) $}\,;\\
\prod_{j=0}^{d_1-r-1} \frac{(d_2+j)!j!}{(r+j)!(d_2-r+j)!}\,, & \text{if\/ $\,\dv = \dmax' +1 $}\,; \\
\infty\,, & \text{otherwise}\,.
\end{cases}
\end{equation}
\end{cor}

\section{Local Unambiguous State Discrimination}\label{Random States}

Pure state discrimination is the task of identifying the state of a quantum system, $\psi_{?}$, given that it is one of a
finite known set of possibilities, $\psiset = \{ \psi_1,\dots,\psi_n \}$. Measurements may be performed on the system, and its
state deduced from the results. If the system is multipartite, we can restrict these measurements to local operations on each
subsystem, coordinated via classical communication. This restriction is abbreviated LOCC, and is particularly important in
quantum information theory, providing the context in which entanglement is a useful resource.

`Unambiguous' state discrimination is the task of obtaining \emph{certain knowledge\/} of the
state~\cite{conclusive1,conclusive2,conclusive3}. When the states in $\psiset$ are nonorthogonal, this certainty when successful
involves a commensurate risk of failure. The risk is significant: educated guesswork is correct more often than an unambiguous
discrimination yields a definite answer. But guesses can be wrong, whereas unambiguous identifications are infallible. It is not
obvious a priori that unambiguous discrimination should be possible at all. If physical states cannot be perfectly distinguished,
why should we expect to separate them with \emph{any\/} probability? The classical analogues of nonorthogonal quantum states are
overlapping probability distributions on phase space, and these cannot be unambiguously distinguished. In this sense, unambiguous
state discrimination is inherently nonclassical.

Linearly independent quantum states can always be globally unambiguously distinguished, but this is not the case locally. (For
example, the Bell states can be perfectly distinguished globally, but cannot be unambiguously distinguished using LOCC.)
Nevertheless, Ji, Cao and Ying~\cite{Ji} showed that any pair of pure states is equally unambiguously distinguishable locally and globally.
Bandyopadhay and one of us~\cite{Bandyo} showed that triplets of quantum states always contain at least one locally identifiable state.
Duan~{\it et al.}~\cite{Duan} showed that any basis of a multipartite Hilbert space, $\H=\bigotimes_j \H_j$ where $\dim\H_j=d_j$,
always contains some locally unambiguously distinguishable subset with $1 + \sum_{j} (d_{j}-1)$ members. They have also presented
a fascinating construction for complete bases of entangled states that are locally unambiguously distinguishable.
Generalizing from these specific results is difficult, as the class of LOCC operations is mathematically complicated.
We aim at the next best thing: a solution for \emph{almost all\/} sets of quantum states.

\bigskip

Every LOCC protocol amounts to some general quantum operation applied to a multipartite Hilbert space $\H$.
Since state discrimination is concerned only with the outcomes of this operation, whose probabilities are described by
a positive operator valued measure (POVM), unambiguous local distinguishability can be defined
in the following way.

A POVM is a set of positive operators $\{E_k\}_k$ which sum to the identity: $\sum_k E_k=1$. POVMs
describe the outcome statistics of general quantum operations. The outcomes are labeled by the
integer $k$. An {\em LOCC decomposable POVM\/} is one which describes an operation on $\H$ that can be implemented by
local operations and classical communication.

\begin{dfn}
The $n$ pure states $\psi_1,\dots,\psi_n$ of the multipartite Hilbert space $\H$ are \emph{unambiguously locally
distinguishable\/} if and only if there exists an LOCC decomposable POVM $\{E_k\}_{k=0,\dots,n}$ on $\H$ with the property that,
\begin{equation}
\langle \psi_{j} | E_k |\psi_{j}\rangle = 0  \;\;\text{ if and only if }\;\; j \neq k\;,
\end{equation}
whenever $k>0$.
\end{dfn}

An outcome $k>0$ unambiguously identifies the state $\psi_{k}$, since no other member of $\psiset$ can produce this result. The
outcome $0$ identifies no state. For any LOCC protocol, every operator $ E_k $ is separable, and can be expressed as a weighted
sum of projectors onto pure product states: $ E_k = \sum_{l} w_{kl} P_{kl}$. The set of all these projectors $\{w_{kl} P_{kl}
\}_{k,l}$ is a POVM describing a separable quantum measurement on $\H$, but not necessarily one that can be implemented using
LOCC. However, for any finite set of product state projectors $\{ P_{l} \}_l$, we can assign nonzero weights $w_{l}'$ such that
the POVM $\{ w_{l}' P_{l}^{\phantom{1}} , 1 - \sum_{l} w_{l}' P_{l}^{\phantom{1}} \}_l$ \emph{does\/} describe an LOCC protocol.
The following condition is then a consequence, first observed by Chefles:

\begin{condition}[Chefles~\cite{Chefles}] \label{c1}
Any\/ $n$ pure states\/ $\psi_1,\dots,\psi_n$ of a multipartite Hilbert space\/ $\H$ are unambiguously locally
distinguishable if and only if there exist\/ $n$ product states\/ $\phi_1,\dots,\phi_n$ of\/ $\H$ with the property that,
\begin{equation}
\langle \phi_{j} | \psi_{k}\rangle = 0   \;\;\text{ if and only if }\;\; j \neq k\;.
\end{equation}
\end{condition}

This condition addresses whether a set of states is unambiguously locally distinguishable or not. It does not indicate the
probability with which this might be achieved, only whether or not this probability is nonzero. This qualitative question is all
that we will consider in this section. We discuss the quantitative problem in Sec.~\ref{open}.

Clearly, for any $\psiset$, searching the set of all product states on $\H$ for states with the necessary properties is far from
trivial. However, with the aid of Theorem~\ref{subspace1} it is simple to prove that the local unambiguous distinguishability of
almost all sets of states can be judged solely by their number in relation to their multipartite structure.

\begin{thm} \label{states}
Let\/ $\H=\bigotimes_j \H_j$ be a multipartite Hilbert space where\/ $\dim\H_j=d_j$.
Then the members of almost all sets of\/ $n$ pure states of\/ $\H$ are locally unambiguously distinguishable,
if and only if
\begin{equation}
n \;\leq\; 1+\sum_j (d_j-1) \;.
\end{equation}
\end{thm}

\begin{proof}
Let $\psiset=\{ \psi_1,\dots,\psi_n \}\subset\P(\H)$. Define $\psiset_1\coloneq\psiset\setminus\{\psi_1\}$, the subset of $\psiset$ including all members but $\psi_1$,
and $\V_1\coloneq\Span\{\ket{\psi}:\psi\in\psiset_1\}$,
the subspace of $\H$ spanned by members of $\psiset_1$. Now let $\V_1^{\perp}$ be the complement of
$\V_1$ in $\H$, i.e. $\H= \V_1^{\phantom{1}} \oplus \V_1^{\perp} $, and define
$\Phi_1\subset\P(\V_1^{\perp})$ to be the set of all product states orthogonal to $\V_1$.

Suppose $n > 1 + \sum_j (d_{j}-1)$. From Theorem \ref{subspace1}, $\Phi_1 = \emptyset$
for almost all choices of $\psiset\subset\P(\H)$. In such cases there is no product state $\phi_1$
with $\langle \phi_1 | \psi_{k}\rangle = 0$ for all $k\neq 1$. Thus
$\psiset$ cannot satisfy Condition~\ref{c1} and is therefore not locally unambiguously distinguishable.

Now suppose $n \leq 1 + \sum_j (d_j-1)$. From Theorem \ref{Wallach}, $\Phi_1 \neq \emptyset$
for all choices of $\psiset$. Now since $\psi_1$ was chosen at random, independently from the members of
$\psiset_1$, and therefore independently from the members of $\Phi_1$, for any product state
$\phi_1 \in \Phi_1$ we almost surely have $\langle \phi_1|\psi_1\rangle \neq 0$, and moreover,
$\langle \phi_1 | \psi_{k}\rangle = 0$ if and only if $k \neq 1$.
By symmetry, similar product states can be found for all members of $\psiset $, which means Condition~\ref{c1}
is satisfied. The members of $\psiset $ are therefore almost surely locally unambiguously distinguishable.
\end{proof}

\subsection{General versus product pure states}

The members of $\psiset $, since they are chosen at random, will be highly entangled in large dimensions. It is interesting to compare this
situation with a more restricted case: sets of randomly chosen \emph{product\/} states. We define a random product state as
follows. For a given Hilbert space $\H = \bigotimes_j \H_{j}$ with $\dim\H_j=d_j$, a random
product state $\phi$ of $\H$ is a tensor product state $|\phi\rangle=\bigotimes_j |\phi_{j}\rangle$ where each state
$\phi_{j}$ is an independently chosen random pure state of $\H_{j}$.

It has been known for some time that almost all sets of $n$ random product states form a (nonorthogonal) unextendible product
basis if and only if $n \geq 1 + \sum_j (d_j-1)$ (see e.g. DiVincenzo~{\it et al.}~\cite{DiVincenzo}). An unextendible product basis is a
set of product states with a span whose complementary subspace contains no product states. The following is then an immediate
consequence of Chefles' condition.

\begin{thm} \label{prods}
Let\/ $\H=\bigotimes_j \H_j$ be a multipartite Hilbert space where\/ $\dim\H_j=d_j$.
Then the members of almost all sets of\/ $n$ \underline{product} pure states of\/ $\H$ are locally unambiguously distinguishable,
if and only if
\begin{equation}
n \;\leq\; 1+\sum_j (d_j-1) \;.
\end{equation}
\end{thm}

Comparing this criterion with that for random general pure states we see something striking: they are identical. Random general
pure states are near-maximally entangled; one might reasonably expect that this makes them harder to distinguish using LOCC. At
least in respect of unambiguous distinguishability, this is not the case.

\subsection{Multiple separated copies}

There may be more than one copy of the system whose state we are trying to identify. Suppose there are $c$ identical
copies of an unknown state $\psi_{?} \in \psiset$. The larger $c$ gets, the easier distinguishing members of $\psiset$ becomes.

The LOCC constraint prohibits joint quantum operations between parties but not between copies, so a party with access to multiple
copies of the same subsystem could measure them all jointly. It is interesting to impose an additional constraint on the parties,
however, and insist that separate operations be performed on every subsystem from every copy -- subsystems from different copies
may not be measured jointly. The copies are then called {\em separated copies}. Equivalently, we extend the scenario of $1$ copy
shared between $m$ parties to that of $c$ copies shared between $cm$ parties, all communicating freely.

This constraint is natural from a practical perspective, especially in a laboratory context. Copies of a system might be
generated and studied at different times. Even if not, whatever technical restriction forbids joint measurements within a copy
could similarly restrict measurements between copies. A relevant example is quantum state tomography, which is almost exclusively
performed on many copies of a system using completely separable measurements \cite{tomography1,tomography2}.

\begin{thm} \label{multiple}
Let\/ $\H=\bigotimes_j \H_j$ be a multipartite Hilbert space with\/ $\dim\H_j=d_j$, and let\/ $c$ be any positive integer.
Then the members of almost all sets of\/ $n$ pure states of\/ $\H$ are locally unambiguously
distinguishable, given $c$ separated copies of the unknown member, if and only if
\begin{equation}
n \;\leq\; 1+c\sum_j (d_j-1) \;.
\end{equation}
\end{thm}

\begin{proof}
Simply consider $\H^{\otimes c}=(\bigotimes_{j=1}^m \H_j)^{\otimes c}$ as the Hilbert space of $cm$ separate parties.
The proof is then exactly the same as that for Theorem~\ref{states}, except Theorem~\ref{subspace1} needs to be replaced by
Corollary~\ref{subspace4}.
\end{proof}

How good are separable measurements at producing unambiguous data? We have seen that above a low threshold size, the members of
almost all sets $\psiset$ are locally unambiguously indistinguishable. With sufficiently many copies, however, the members of any
finitely large $\psiset$ can be unambiguously distinguished. The minimum number of copies this requires gives a measure of
\emph{how\/} indistinguishable $\psiset$ is. Contrasting the minima for separable and global measurements shows how poor the
former are at obtaining that `quintessentially quantum' unambiguous data. As the dimension of the measured system increases,
separable measurements consume exponentially more copies.

\begin{cor}
Let\/ $\H=\bigotimes_j \H_j$ be a multipartite Hilbert space with\/ $\dim\H_j=d_j$.
Then for almost all sets of\/ $n$ pure states of\/ $\H$,
\begin{equation}
c \;\geq\;  \frac{n-1}{\sum_j (d_j-1)}
\end{equation}
identical copies are required to locally unambiguously identify an unknown member, if these copies are separated.
\end{cor}

What does this imply about the feasibility of unambiguous discrimination on higher dimensional multipartite systems? Suppose we
grow the number of subsystems, $m$. For any fixed number of copies, the number of random states that can be unambiguously
distinguished globally grows exponentially with $m$. But far from being fixed, the number of copies required to emulate these
global measurements \emph{separably\/} also grows exponentially. A linearly independent set of $m$-qubit states, for example, has
up to $2^m$ members. To unambiguously distinguish this many states using single qubit measurements requires
$\big\lceil\frac{2^m-1}{m}\big\rceil$ copies of the system. Globally, one is always enough!

Of course with many copies of a quantum system, a great deal can be learned about its state without using unambiguous protocols.
As measurement results accumulate, analysis of $\psiset$ will reveal most states becoming increasingly unlikely and a select
few candidates becoming highly probable. What will remain elusive is certainty. In theory (if admittedly not in practice)
absolute certainty is obtainable by global measurements, and exponentially harder to reach with separable operations.

\section{Discussion}\label{open}

Almost all subspaces of multipartite Hilbert spaces are completely entangled, provided they are small enough. We have precisely
quantified `small enough', and it is very large (\ref{bound}). For any $m$-qubit system, almost all subspaces of dimension
$2^m-m-1$ or less are completely entangled. Random and entangled subspaces are useful in a number of contexts, and we hope our result
will find application elsewhere. It should certainly make them easier to find! One application we have pursued ourselves,
obtaining a simple rule governing the local unambiguous distinguishability of almost all sets of quantum states.

We have only discussed pure states, but our results have implications for random mixed states as well. Hayden \emph{et
al}.~\cite{Hayden1} showed that general mixed states of rank equal to the span of a `maximally entangled' random subspace, have
near maximal entanglement of formation but near zero distillable entanglement, making them nearly bound entangled. Cubitt
\emph{et al}.~\cite{Cubitt} observe their constructions can be used to created similar mixed states of much greater rank with
high Schmidt measure. Our results suggest this property of mixed states is generic.

Of the many questions left open, the most obvious concerns quantification. Precisely \emph{how\/} entangled are the states in a
completely entangled random subspace? The Schmidt rank provides one measure in bipartite situations. What of other measures, such
as the entanglement of formation, or of distillation? Whereas the maximally entangled subspaces of Hayden \emph{et al}. contain
only states with high entropy of entanglement, our much larger subspaces sacrifice this property as they grow. The largest
completely entangled subspaces contain states that are only slightly entangled. Thanks to Corollary~\ref{Schmidt}, we can chart
the stepwise descent of random subspaces' Schmidt rank as their dimension increases. It would be interesting to see how other
entanglement measures behaved.

Distinguishability can be quantified as well. If a set of random states can be locally unambiguously discriminated, with what
probability of success? If not, how many copies would provide a better than even odds chance? Unfortunately, we still do not have
good bounds on the probability with which \emph{global\/} unambiguous discrimination can be performed on random states, so
answers to these local questions remain out of reach. Good progress has been made on the optimal global distinguishability of
random states \cite{Montanaro}, however, so both the unambiguous global problem and the optimal local problems are tantalizing.

We can make one locally unambiguous observation, however. If the members of a set of pure states can be locally unambiguously
discriminated with nonzero probability, then they can certainly be locally unambiguously discriminated with some finite,
nonnegligible probability. The probability overlap of a random state $\psi_{i}$ with some arbitrary product state that identifies
it, $\phi_{i}$, will have mean $\overline{\,|\braket{\phi_{i}}{\psi_{i}}|^2}=1/D$, where $D$ is the dimension of the overall
system. (To see this, consider writing the random state in an independently fixed product basis.) As $D$ becomes large,
concentration of measure effects transform this average into a very reliable estimate. Thus selecting an arbitrary set of product
states, $\{\phi_{i}\}$, satisfying Chefles' Condition~\ref{c1}, we can build an LOCC protocol around them that stands a finite
chance of success: $\{ w_{i} |\phi_{i}\rangle \langle \phi_{i}|, 1 - \sum_{i} w_{i} |\phi_{i}\rangle \langle \phi_{i}| \}_i$. The
weights $w_{i}$ must be considered because the measurement must be LOCC implementable, but these can be quite large. For instance
if for all $i$, $w_{i} = \frac{1}{n}$, the POVM is automatically LOCC decomposable irrespective of $\{\phi_{i}\}$. Such a
protocol simply guesses the state's identity in advance, then performs binary local measurements to confirm or refute this
hypothesis. This yields a success probability of about $\frac{1}{Dn}$. There will be much better choices of measurement, of
course, but even arbitrary successful protocols succeed a noticeable fraction of the time.

If $n$ random product pure states are locally unambiguously distinguishable then so are $n$ random general pure states, and vice
versa. Entanglement doesn't matter for determining whether or not the states are distinguishable, but perhaps the two sets can be
distinguished with different probabilities of success. If the probability for general states were lower, we might recover some of
our intuition about entanglement. We cannot judge this question now, but observe that \emph{arbitrary\/} successful protocols
show no sign of favouring separable states. This is not conclusive evidence one way or the other, but it reinforces that there is
no evidence for random highly entangled states being harder to locally unambiguously distinguish than random product states.

\section{Acknowledgements}

We would like to thank Somshubhro Bandyopadhyay, Debbie Leung, and John Watrous for useful discussions. Research at the Perimeter
Institute for Theoretical Physics is supported in part by the Government of Canada through NSERC and by the province of Ontario
through MRI. AJS is supported by ARC and the State of Queensland, and thanks the Perimeter Institute for their hospitality.

\end{document}